\documentclass[twocolumn,showpacs,showkeys,preprintnumbers,amsmath,amssymb,prb,floatfix]{revtex4-1}

\usepackage{amsfonts}
\usepackage{amssymb}
\usepackage{graphicx}
\usepackage{dcolumn}
\usepackage{bm}
\begin{document}

\title{Spin-orbital Texture in Topological Insulators}

\author{Haijun Zhang$^1$, Chao-Xing Liu$^2$  \& Shou-Cheng Zhang$^{1}$}

\affiliation{
$^1$ Department of Physics, McCullough Building, Stanford University, Stanford, CA 94305-4045\\
$^2$ Department of Physics, The Pennsylvania State University, University Park, Pennsylvania 16802-6300
}

\date{\today}

\begin{abstract}
{\bf Relativistic spin-orbit coupling plays an essential role in the field of topological insulators and
quantum spintronics. It gives rise to the topological non-trivial band structure and enables electric
manipulation of the spin degree of freedom. Because of the spin-orbit coupling, rich spin-orbital
coupled textures can exist both in momentum and in real space. For three dimensional topological
insulators in the Bi$_2$Se$_3$ family, topological surface states with p$_z$ orbitals have a left-handed
spin texture for the upper Dirac cone and a right-handed spin texture for the lower Dirac cone.
In this work, we predict a new form of the spin-orbital texture associated with the p$_x$ and
p$_y$ orbitals. For the upper Dirac cone, a left-handed (right-handed) spin texture is coupled to
the ``radial'' (``tangential'') orbital texture, whereas for the lower Dirac cone, the coupling of
spin and orbital textures is the exact opposite. The ``tangential'' (``radial'') orbital texture is dominant for the upper
(lower) Dirac cone, leading to the right-handed spin texture for the in-plane orbitals of both the upper and lower Dirac cones.
A spin-resovled and photon polarized angle-resolved photoemission
spectroscopy experiment is proposed to observe this novel spin-orbital texture.}
\end{abstract}

\maketitle

\section{introduction}
Three-dimensional topological insulators (TIs) are new states of quantum matter with helical gapless surface states consisting of odd number of Dirac cones inside the bulk band gap protected by time-reversal symmetry (TRS).\cite{qi2010a,Moore2010,Hasan2010,Qi2011} The underlying physical origin of the topological property of TIs is the strong spin-orbit coupling (SOC), which plays a similar role as the Lorentz force in the Quantum Hall state. Due to the SOC interaction, the spin and momentum are locked to each other, forming a spin texture in the momentum space for the surface states of TIs\cite{zhang2009,Zhang2010,Yazyev2010}. The spin texture has been directly observed in the spin-resolved angle-resolved photon emission spectroscopy (spin-resolved ARPES)\cite{Hsieh2009,Souma2011,xu2011,pan2011,Jozwiak2011}. The spin texture gives rise to a non-trivial Berry phase for the topological surface states and suppresses the backscatterings under TRS, leading to possible device applications in spintronics.

Besides the spin texture, it has also been shown recently that the atomic $\mathbf{p}$ orbitals of the Bi$_2$Se$_3$ family of topological
insulators form a pattern in the momentum space, dubbed as the orbital texture, for the topological surface states.\cite{Park2012,Cao2012}
In this work, we predict a coupled spin-orbital texture for the topological surface states. Based on both the effective {\bf k$\cdot$ p} theory and {\it ab-initio} calculations, we find, besides the usual locking between the electron spin and the crystal momentum, the spin texture is
also locked to the atomic orbital texture, which is dubbed as ``spin-orbital texture''. We show that $p_z$ orbitals have left-handed spin texture for the upper Dirac cone and right-handed spin texture for the lower Dirac cone, sharing the same feature as the total spin texture of the surface states. In contrast, the in-plane orbitals ($p_x$ and $p_y$ orbitals) reveal more intriguing features: for the upper Dirac cone of surface states, a ``radial'' orbital texture is coupled to a left-handed spin texture and a ``tangential'' orbital texture is coupled to a right-handed spin texture. For the lower Dirac cone, the coupling between spin and orbital textures is exactly opposite. An electron spin-resolved and
photon polarized ARPES experiment is proposed to observe this novel spin-orbital texture of the surface states of TIs.

\begin{figure}
   \begin{center}
      \includegraphics[width=1.3in,angle=-90]{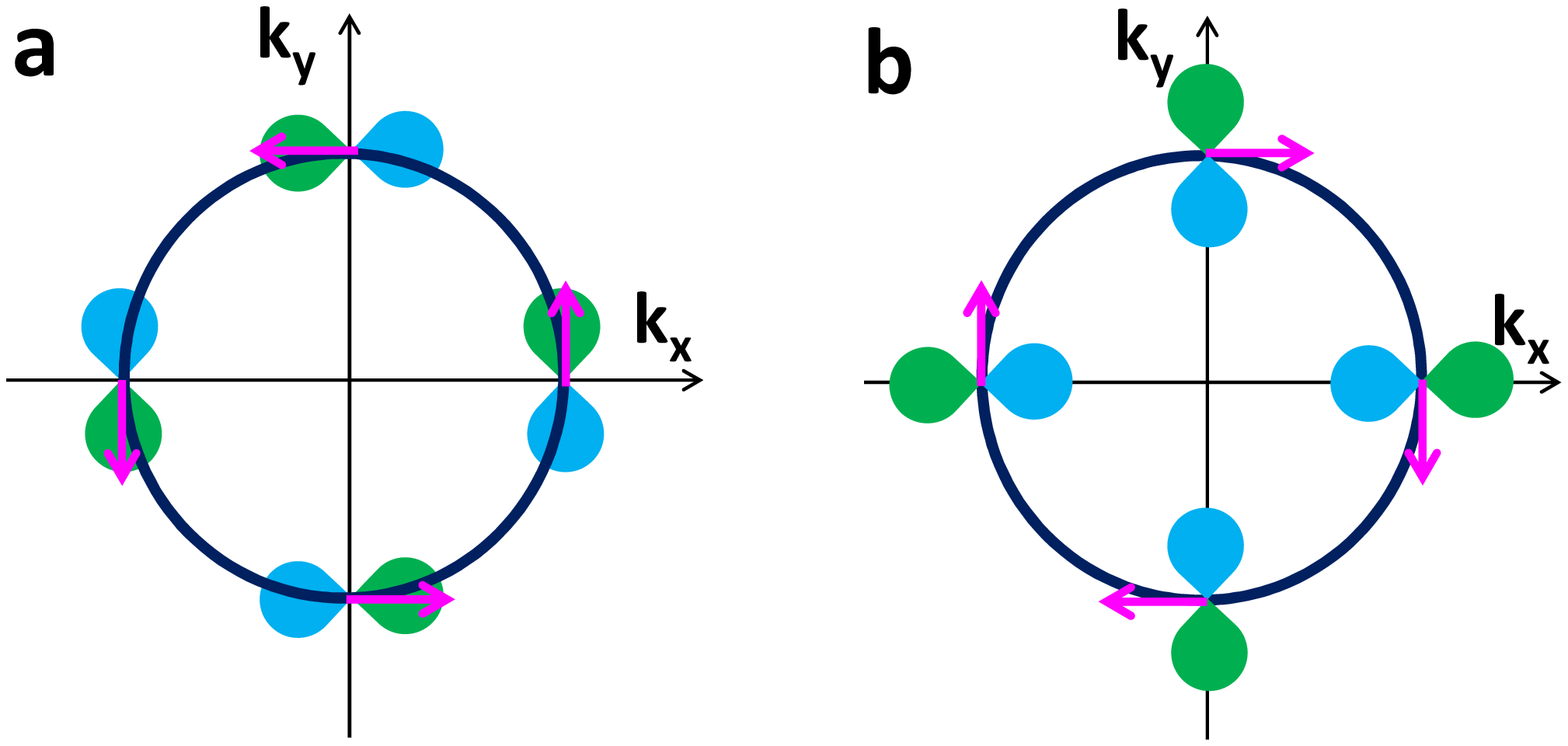}
    \end{center}
    \caption{ (color online)
     {a}, { b}, The tangential orbital texture with the right-handed helical spin texture {(a)} and the radial orbital texture with the left-handed helical spin texture {(b)} for the upper Dirac cone.}
    \label{fig1}
\end{figure}

\begin{figure*}
   \begin{center}
      \includegraphics[width=4.5in,angle=-90]{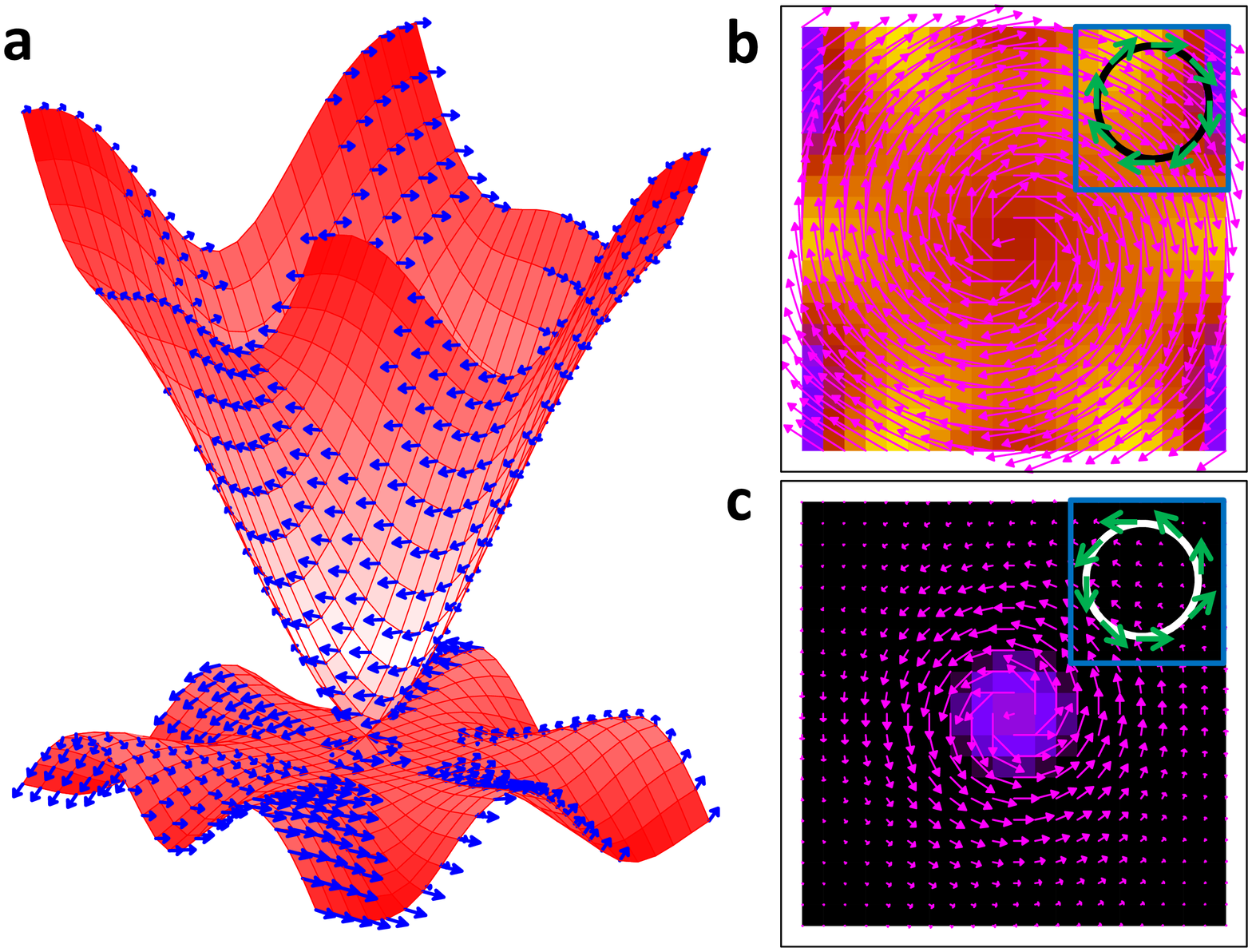}
    \end{center}
    \caption{ (color online)
     { a}, The Dirac cone of Bi$_2$Se$_3$ on the surface with the normal direction [0001] with the spin texture marked by blue arrows. { b}, { c}, The projection of $p_z$ orbital and the related in-plane spin texture for upper { (b)} and lower {\bf (c)} Dirac cones.  More red means more $p_z$ character. The red arrows represent the in-plane spin texture related to the $p_z$ orbitals. The insets are the schematics of the spin texture marked by green arrows.}
    \label{fig2}
\end{figure*}

\begin{figure*}
   \begin{center}
      \includegraphics[width=4.5in,angle=-90]{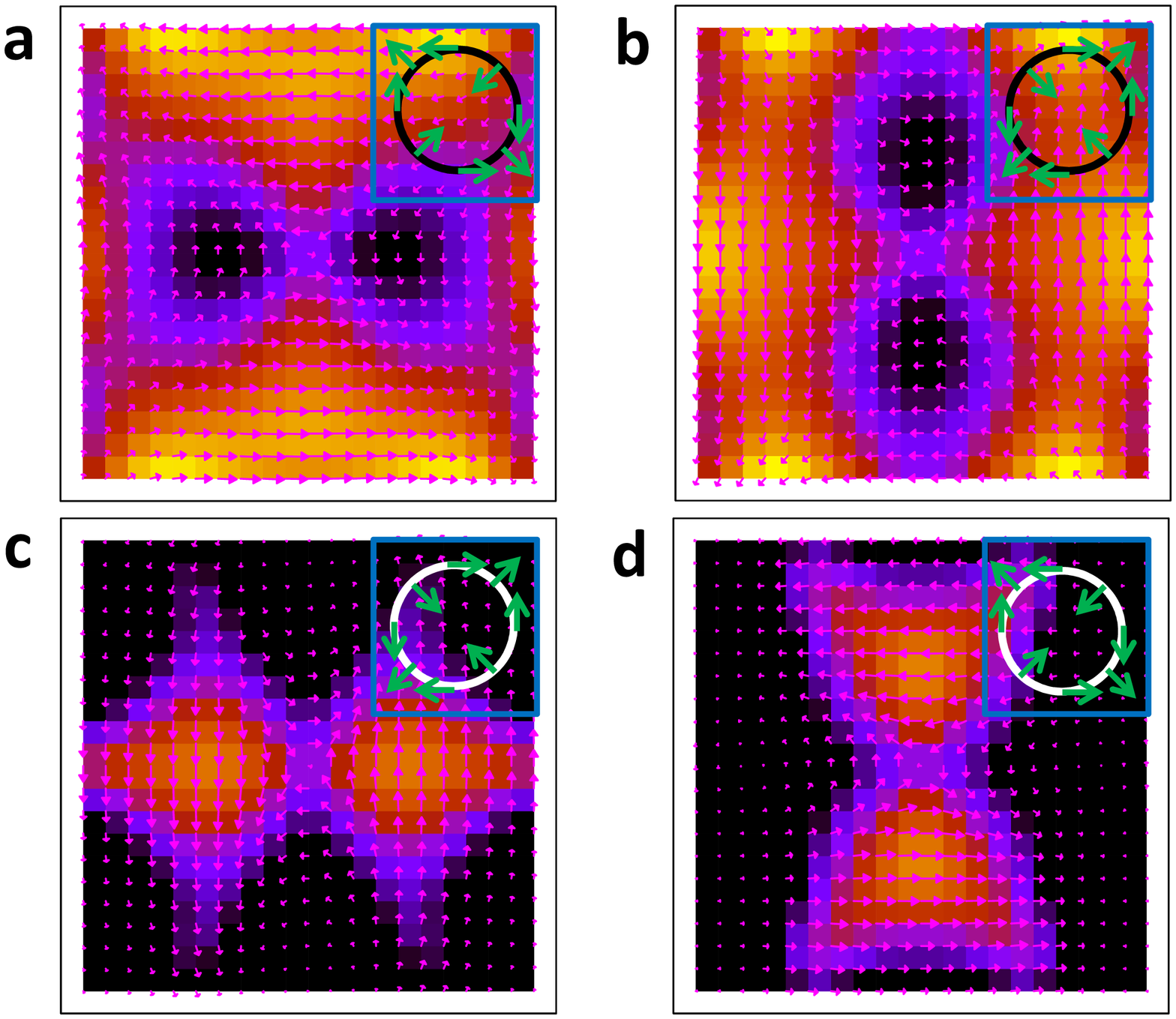}
    \end{center}
    \caption{ (color online)
     { a},{ b},{ c},{ d}, The $p_x$ projection on the states of upper { (a)} and lower { (c)} Dirac cones, and the $p_y$ projection of upper { (b)} and lower { (d)} Dirac cones. More red means more $p_x$ character in {(a)} and {(c)}, and  more red means more $p_y$ character in {(b)} and {(d)}. The red arrows indicate the in-plane spin texture related to the orbitals. The insets are the schematics of the spin texture.
 }
    \label{fig3}
\end{figure*}

\begin{figure*}
   \begin{center}
      \includegraphics[width=4.5in,angle=-90]{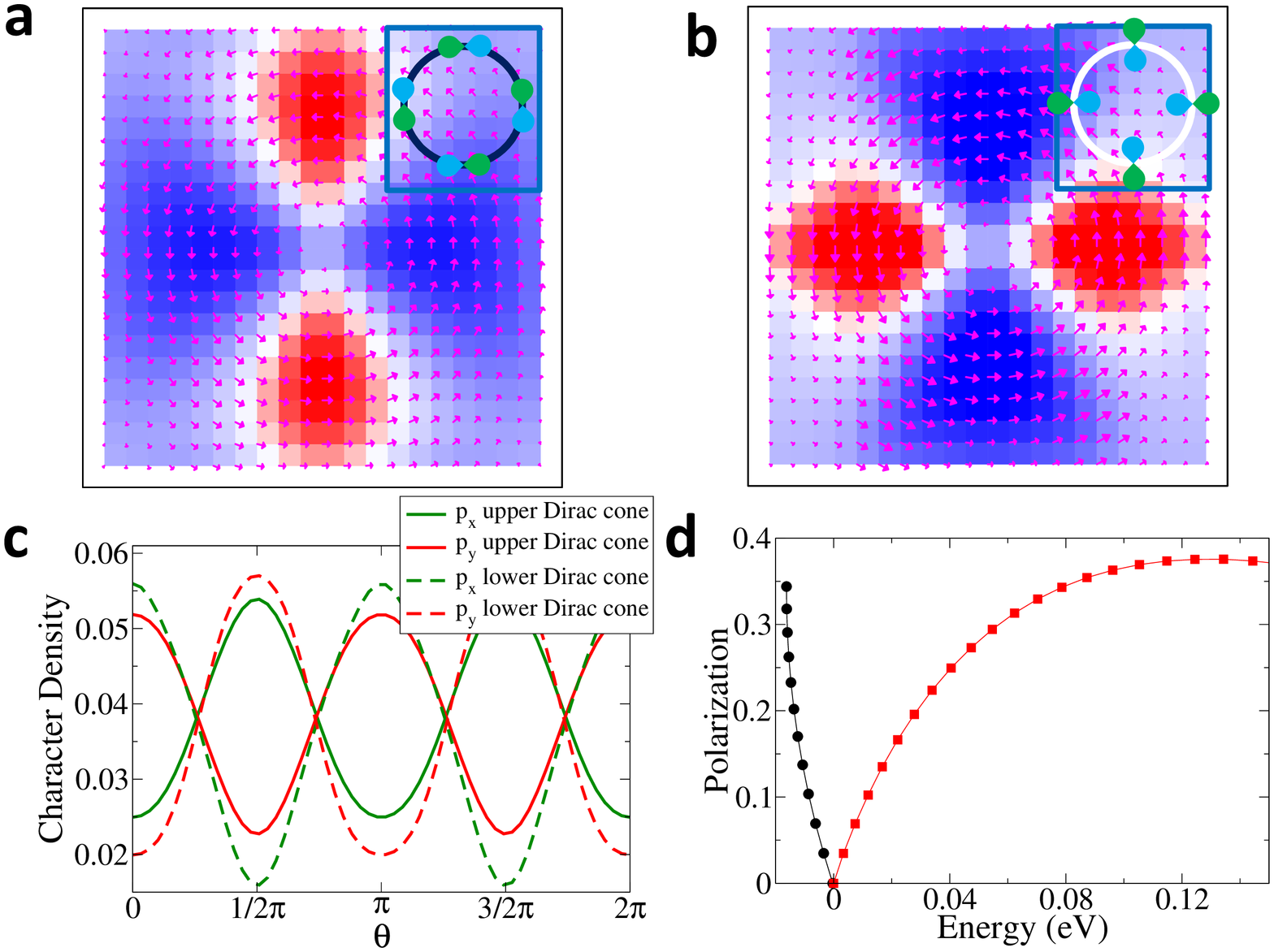}
    \end{center}
    \caption{ (color online)
      { a},{ b}, The tangential orbital texture with the related in-plane spin texture for the upper Dirac cone { (a)} and the radial orbital texture for the lower { (b)} from {\it ab-initio} calculations. More red means more $p_x$ character, and more blue means more $p_y$ character. The red arrows represent the total in-plane spin texture related to $p_x$ and $p_y$ orbitals. { c}, The $p_x$ and $p_y$ projections onto upper and lower Dirac cones. The solid curves are the $p_x$ and $p_y$ projections for the upper Dirac cone at energy level $0.10$eV, and the dashed for the lower Dirac cone at the energy $-0.07$eV. The basic feature is the $\pi$ period which exactly agrees with the prediction by the effective model. { d}, Orbital polarization $P_{p_x}$. The energy of Dirac point is shifted to be zero. The positive value of the orbital polarization represents the radial orbital texture, and the negative value represents the tangential orbital texture. In order to plot more visually, the $P_{p_x}$ for the upper Dirac cone is inversed, marked by the red color. $P_{p_x}$ is exact zero at Dirac point, which indicates the transition point between the tangential and  radial orbital textures.
 }
    \label{fig4}
\end{figure*}

\section{Effective theory of the spin-orbital texture}
The surface states of TIs are described by the Dirac type of effective Hamiltonian\cite{zhang2009,liu2010}
\begin{eqnarray}
H_{\rm
surf}(k_x,k_y)=\hbar v_f\left(\sigma^xk_y-\sigma^yk_x\right),
\label{eq:sur_Heff}
\end{eqnarray}
with the Fermi velocity $v_f$ and Pauli matrix $\sigma$. The salient feature of this effective Hamiltonian is the ``spin-momentum locking'', which means for a fixed momentum {\bf k}, the ``spin'', denoted by the Pauli matrix $\sigma$, has a fixed direction for the eigenstate of the Hamiltonian. Since the ``spin'' is always perpendicular to the momentum, we can introduce a helicity operator, defined as $\hat{h}=\frac{1}{k}\hat{z}\cdot(\vec{k}\times\vec{\sigma})$ which commutates with the Hamiltonian, to determine the handness of the ``spin'' texture. For the upper Dirac cone of surface states, the helical operator $\hat{h}=-1$, leading to a left-handed ``spin'' texture in the momentum space while for the lower Dirac cone, $\hat{h}=1$ yields a a right-handed ``spin'' texture. However, one should note that here the ``spin'' is not the real spin, but the total angular momentum $\vec{J}=\vec{S}+\vec{L}$, which is a combination of the real spin $\vec{S}$ and the orbital angular momentum $\vec{L}$ due to SOC. Consequently, the basis of the surface effective Hamiltonian (\ref{eq:sur_Heff}) are denoted as $|\Psi_{J_z=\pm\frac{1}{2}}\rangle$ with the lower indices $\pm\frac{1}{2}$ representing the total angular momentum along $z$ direction. In order to understand what is the texture for the real spin $\vec{S}$, it is necessary to write down the explicit form of the basis wavefunction $|\Psi_{\pm\frac{1}{2}}\rangle$.

The form of the basis $|\Psi_{\pm\frac{1}{2}}\rangle$ can be constructed by symmetry considerations. Generally the basis $|\Psi_{\pm\frac{1}{2}}\rangle$ depends on the momentum {\bf k} and we can expand it up to the first order in {\bf k} as $|\Psi_{\pm\frac{1}{2}}\rangle=|\Psi^{(0)}_{\pm\frac{1}{2}}\rangle+|\Psi^{(1)}_{\pm\frac{1}{2}}\rangle$. Here we are only interested in the $\mathbf{p}$ orbitals of Bi and Se atoms in the topological insulator Bi$_2$Se$_3$ and can decompose the zeroth-order wavefunction as
\begin{eqnarray}
	&&|\Psi^{(0)}_{\pm\frac{1}{2}}\rangle=\sum_\alpha \left[u_{0,\alpha}|\alpha,p_z,\uparrow(\downarrow)\rangle+v_{0,\alpha}|\alpha,p_{\pm},\downarrow(\uparrow)\rangle \right]
	\label{eq:surwf0}
\end{eqnarray}
and the first-order wavefunction as
\begin{eqnarray}\label{eq:surwf1}
\begin{split}
|\Psi^{(1)}_{\pm\frac{1}{2}}\rangle=
\sum_\alpha[&\pm k_\pm(iu_{1,\alpha}|\alpha,p_\mp,\uparrow(\downarrow)\rangle+iv_{1,\alpha}|\alpha,p_z,\downarrow(\uparrow)\rangle)\\
&\mp iw_{1,\alpha}k_\mp|\alpha,p_\pm,\uparrow(\downarrow)\rangle]
\end{split}
\end{eqnarray}
where $k_\pm=k_x\pm i k_y$, $|\uparrow\rangle$ and $|\downarrow\rangle$ denote the spin, $|p_z\rangle$ and $|p_\pm\rangle=\mp\frac{1}{\sqrt{2}}\left( |p_x\rangle\pm i|p_y\rangle \right)$ denote different $p$ orbitals, and $\alpha$ denotes indices other than the spin and orbital, such as atom indices. Here $u_{0(1),\alpha}$, $v_{0(1),\alpha}$ and $w_{1,\alpha}$ are material-dependent parameters. By comparing with the {\it ab-initio} calculations, we find that we can take them to be real. $|\Psi_{\frac{1}{2}}\rangle$ and $|\Psi_{-\frac{1}{2}}\rangle$ are related to each other by TRS. The expressions of the basis (\ref{eq:surwf0}) and (\ref{eq:surwf1}) can be substituted into the eigen wavefunctions of the Hamiltonian (\ref{eq:sur_Heff}),
$|\Phi_\pm\rangle=\frac{1}{\sqrt{2}}\left[ \pm ie^{-i\theta_k}|\Psi_{\frac{1}{2}}\rangle+ |\Psi_{-\frac{1}{2}}\rangle \right]$,
yielding the following forms of the wavefunctions
\begin{eqnarray}
\begin{split}
	|\Phi_{+}\rangle=\sum_\alpha[& (u_{0,\alpha}-v_{1,\alpha}k)|\alpha, p_z, \uparrow_\theta\rangle\\
    & -\frac{i}{\sqrt{2}}(v_{0,\alpha}-u_{1,\alpha}k-w_{1,\alpha}k)|\alpha, p_{r}, \uparrow_\theta\rangle \\
	 &+ \frac{1}{\sqrt{2}} (v_{0,\alpha}-u_{1,\alpha}k+w_{1,\alpha}k) |\alpha,p_t, \downarrow_\theta\rangle]\label{eq:eigenwf1}\\
\end{split}
\\
\begin{split}
	|\Phi_{-}\rangle=\sum_\alpha[ & (u_{0,\alpha}+v_{1,\alpha}k)|\alpha, p_z,\downarrow_\theta\rangle \\
   &+\frac{i}{\sqrt{2}} (v_{0,\alpha}+u_{1,\alpha}k+w_{1,\alpha}k)|\alpha, p_r, \downarrow_\theta\rangle\\
	&-\frac{1}{\sqrt{2}}(v_{0,\alpha}+u_{1,\alpha}k-w_{1,\alpha}k)|\alpha, p_t, \uparrow_\theta\rangle]
	\label{eq:eigenwf2}.
\end{split}
\end{eqnarray}
Here $|\uparrow_\theta(\downarrow_\theta)\rangle=\frac{1}{\sqrt{2}}( +(-) ie^{-i\theta_k}|\uparrow\rangle +|\downarrow\rangle)$ stands for the left-handed (right-handed) helical spin texture and $|p_r\rangle=\cos\theta_k|p_x\rangle+\sin\theta_k|p_y\rangle$, $|p_t\rangle=-\sin\theta_k|p_x\rangle+\cos\theta_k|p_y\rangle$ are the radial and tangential orbital textures, as shown in Fig.~1a,b, respectively. From the expressions (\ref{eq:eigenwf1}) and (\ref{eq:eigenwf2}), we can clearly see that $|p_z\rangle$ orbital is coupled to the left-handed spin texture $|\uparrow_\theta\rangle$ for the upper Dirac cone and the right-handed spin texture $|\downarrow_\theta\rangle$ for the lower Dirac cone. Furthermore, for the upper Dirac cone, the radial orbital texture $|p_r\rangle$ is always coupled to the left-handed spin texture $|\uparrow_\theta\rangle$ and the tangential orbital texture $|p_t\rangle$ is always coupled to the right-handed spin texture $|\downarrow_\theta\rangle$. The situation is exactly opposite for the lower Dirac cone.

The expressions (\ref{eq:eigenwf1}) and (\ref{eq:eigenwf2}) are the main analytical results of this paper, which show explicitly the spin-orbital texture. To confirm our analytical results, {\it ab-initio} method is adopted to calculate the projection of surface states on the spin and orbital basis, defined by the quantity
\begin{eqnarray}
	D^\pm_{i,\eta}= \langle\Phi_\pm|(|p_i\rangle\langle p_i|\otimes s_\eta)|\Phi_\pm\rangle,
	\label{eq:surProj}
\end{eqnarray}
where $p_i=p_x,p_y,p_z$ for the three $\mathbf{p}$ orbitals, $s_0={\bf 1}_{2\times2}$ denotes the charge part and $s_{x,y,z}$ denote the three Pauli matrices for the spin. In the following, we will compare the analytical calculation of the intensity $D^\pm_{i,\eta}$ with {\it ab-initio} calculations.

\section{Spin-orbital texture from ab-initio calculations}

The Vienna Ab-initio Simulation Package (VASP)\cite{kresse1993,kresse1999} is employed to carry out {\it ab-initio} calculations with the framework of the Perdew-Burke-Ernzerhof-type (PBE)\cite{perdew1996} generalized gradient approximation (GGA) of density functional theory\cite{hohenberg1964}. Projector augmented wave (PAW) pseudo-potentials are used for all of the calculations in this work\cite{Blochl1994}. 10$\times$10$\times$10 and 10$\times$10$\times$1 are used for \textbf{k}-grid of bulk and free-standing calculations, respectively. The kinetic energy cutoff is fixed to 450eV. 6 quintuple layers (QLs) are fixed in the supercell for free-standing calculations, and the thickness of vacuum is taken to be 50${\AA}$. The lattice constant and the atomic parameters are directly obtained from experiments. SOC is included with the non self-consistent calculation. In order to compare with the result of ARPES experiments, the projections of all the orbitals are only for the first Se and Bi atoms on the top surface of the free-standing model.

The surface states of Bi$_2$Se$_3$ consist of a single Dirac cone at $\Gamma$ point on one surface inside the large bulk band gap ($\sim$ 0.3eV)\cite{zhang2009,Xia2009}, which provides an ideal material to study the coupling of spin and orbital textures of surface states. As the starting point, we compare the bulk band structure of Bi$_2$Se$_3$ with the previous calculation\cite{zhang2009} and find good agreements. The surface states are obtained from the calculation of a free-standing structure with the normal direction [0001], as shown in Fig.~2a. The blue arrows represent the spin texture, where the spin is mainly lying in plane near the Dirac point. The spin texture is left-handed for the upper Dirac cone and right-handed for the lower one, the same as the total angular momentum texture. To understand the underlying physics, we calculate the spin texture for different atomic orbitals. For $p_z$ orbitals, a left-handed helical spin texture is found for the upper Dirac cone and a right-handed texture for the lower Dirac cone, as shown in Fig.~2b,c. The schematic of the spin texture are shown in the inset.
Here the background color indicates the projection of $p_z$ orbitals, which is isotropic, and the red arrows represent the corresponding in-plane spin texture. The spin texture of $p_z$ orbitals can be reproduced with the expressions $[D^\pm_{p_z,\sigma_x},D^\pm_{p_z,\sigma_y},D^\pm_{p_z,\sigma_z}]=\pm\sum_\alpha(u_{0,\alpha}\mp v_{1,\alpha}k)^2[\sin\theta_k,-\cos\theta_k,0]$ with `$\pm$' for the upper and lower Dirac cone and $\vec{k}=(k,\theta_k)$ in the polar coordinate.

The spin textures for in-plane orbitals are shown in Fig.~3a,b for the upper Dirac cone and in Fig.~3c,d for the lower Dirac cone, respectively. We find that the associated spins for $p_x$ and $p_y$ orbitals don't rotate clockwise or anti-clockwise around the Dirac point as in the case of $p_z$ orbitals, but instead, they take the form
\begin{eqnarray}
	&&[D^\pm_{p_{x},x},D^\pm_{p_{x},y},D^\pm_{p_{x},z}]=\mp\sum_\alpha \frac{v^2_{0,\alpha}}{2} [\sin\theta_k,\cos\theta_k,0]\label{eq:Spintexture_px}\\
	&&[D^\pm_{p_y,x},D^\pm_{p_y,y},D^\pm_{p_y,z}]=\pm\sum_\alpha \frac{v^2_{0,\alpha}}{2} [\sin\theta_k,\cos\theta_k,0]\label{eq:Spintexture_py}
\end{eqnarray}
for small $k$ around the $\Gamma$ point. The corresponding spin textures are shown schematically in the inset of Fig.~3a,c for $p_x$ orbitals in upper and lower Dirac cones and in the inset of Fig.~3b,d for $p_y$ orbitals. Unlike $p_z$ orbitals, the amplitude of $p_x$ and $p_y$ orbitals for the surface states is not isotropic, but has $2\theta_k$ angular dependence around the Fermi surface, as shown in Fig.~4c. We may take the difference of the amplitudes between $p_x$ and $p_y$ orbitals, as shown by colors in Fig.~4a,b. Here more red means more $p_x$ character, and more blue means more $p_y$ character. The angular dependence indicates a tangential orbital texture for the upper Dirac cone and a radial orbital texture for the lower Dirac cone, as schematically shown by the inset of Fig.~4a,b, respectively. This orbital texture was experimentally observed recently\cite{Cao2012}.
Furthermore, we also plot the total spin textures for in-plane orbitals on the same figure, which show a right-handed texture for both upper and lower Dirac cones. All these salient feature can be understood by the wavefunctions (\ref{eq:eigenwf1}) and (\ref{eq:eigenwf2}). For the upper Dirac cone, although both $|\uparrow_\theta\rangle|p_{r}\rangle$ and $|\downarrow_\theta\rangle|p_{t}\rangle$ terms exist in the wavefunction (\ref{eq:eigenwf1}), their associated coefficients are unequal. When $\sum_\alpha(v_{0,\alpha}-u_{1,\alpha}k+w_{1,\alpha}k)^2>\sum_\alpha(v_{0,\alpha}-u_{1,\alpha}k-w_{1,\alpha}k)^2$, $|\downarrow_\theta\rangle|p_{t}\rangle$ term dominates over $|\uparrow_\theta\rangle|p_{r}\rangle$ term, dominantly giving a tangential orbital texture coupled to a right-handed spin texture. Similarly, for the lower Dirac cone, when $\sum_\alpha(v_{0,\alpha}+u_{1,\alpha}k+w_{1,\alpha}k)^2>\sum_\alpha(v_{0,\alpha}+u_{1,\alpha}k-w_{1,\alpha}k)^2$, $|\downarrow_\theta\rangle|p_{r}\rangle$ term in the wavefunction (\ref{eq:eigenwf2}) is dominant, yielding a radial orbital texture coupled to a right-handed spin texture. The difference between $p_x$ and $p_y$ orbitals can be calculated directly as $D^\pm_{p_x,0}-D^\pm_{p_y,0}=\mp2\cos2\theta_k\sum_\alpha\left[ (v_{0,\alpha}\mp ku_{1,\alpha})kw_{1,\alpha} \right]$, which indeed shows a $2\theta_k$ angular dependence, and the total spin textures for in-plane orbitals can be obtained as $[D^\pm_{x},D^\pm_{y}]=[D^\pm_{p_x,x}+D^\pm_{p_y,x},D^\pm_{p_x,y}+D^\pm_{p_y,y}]=4[-\sin\theta_k,\cos\theta_k]\sum_\alpha (v_{0,\alpha}\mp ku_{1,\alpha})kw_{1,\alpha}$, which shows a right-handed spin texture when $\sum_\alpha(v_{0,\alpha}\mp ku_{1,\alpha})kw_{1,\alpha}>0$. Especially, if $k$ gets close zero, both the total spin texture $[D^\pm_{x},D^\pm_{y}]$ of the in-plane orbitals and the difference between $p_x$ and $p_y$ orbitals $D^\pm_{p_x,0}-D^\pm_{p_y,0}$ approaches zero, also as shown in Fig.~4a,b.

Therefore, there is a transition from a tangential orbital texture in the upper Dirac cone to a radial orbital texture in the lower Dirac cone, switching exactly at the Dirac point. To quantitatively describe this transition, we introduce a polarization quantity
\begin{equation}\label{polarization}
    P_{p_x}(\pm)=\frac{D_{p_x,0}(\pm,\theta=0)-D_{p_x,0}(\pm,\theta=90)}{D_{p_x,0}(\pm,\theta=0)+D_{p_x,0}(\pm,\theta=90)}
\end{equation}
with `$\pm$' for upper and lower Dirac cones. The plot of $P_{p_x}(\pm)$ is shown in Fig.~4d where the energy level of the Dirac point is shifted to zero. In order to show the plot more visually, we reverse the value of the $P_{p_x}(+)$ for the upper Dirac cone plotted with the red. The feature of $P_{p_x}(\pm)$ undoubtedly indicates that the state of the lower Dirac cone forms a radial orbital texture, and the state of the upper Dirac cone forms a tangential orbital texture. The Dirac point is shown to be the exact transition point from the tangential to radial orbital texture. This is exactly the behavior observed in a recent experiment and explained within the first principle calculations\cite{Cao2012}. The numerical results fit well to the analytical calculation, with the expression
 \begin{equation}\label{polarization1}
 P_{p_x}(\pm)=\mp\frac{2\sum_\alpha (v_{0,\alpha}\mp Eu_{1,\alpha}/\hbar v_f)Ew_{1,\alpha}/\hbar v_f}{\sum_\alpha\left[ (v_{0,\alpha}\mp Eu_{1,\alpha}/\hbar v_f)^2+E^2w_{1,\alpha}^2/\hbar^2v_f^2\right]}
 \end{equation}
 with the energy $E$. For small $E$ around Dirac point, $P_{p_x}(\pm)\propto \mp\frac{\sum_\alpha v_{0,\alpha}w_{1,\alpha}}{\sum_\alpha(v_{0,\alpha})^2}\frac{2E}{\hbar v_f}$ shows the linear dependence on energy, as found in Fig.~4d.

Although in-plane orbitals show different spin textures compared to $p_z$ orbitals, we stress the $p_z$ orbitals ($50\%$) dominate the states near the Dirac point with $p_x$ and $p_y$ only around $30\%$. Therefore, the spin texture for the whole states show left-handed for the upper Dirac cone and right-handed for the lower Dirac cone, the same as that of $p_z$ orbitals, as well as the total angular momentum texture.

%
%

\section{Discussion}

In order to detect the spin texture of electrons, the spin-resolved ARPES technology has been developed by taking advantage of spin-dependent scatting processes and precisely measuring the magnitude of the asymmetry in the spin-dependent intensity with perfect spin-polarimeters.\cite{Jozwiak2011} The non-trivial spin texture of surface states of TIs has been clearly observed by experiments.\cite{Hsieh2009,Souma2011,xu2011,pan2011,Jozwiak2011} In addition, the orbital character can be detected through the photon polarization selection rules\cite{Damascelli2003} based on the symmetry analysis. With this technology, the orbital texture of surface states of Bi$_2$Se$_3$ was reported recently by a polarized ARPES experiment.\cite{Cao2012} Therefore, it is possible to combine these two technologies together in an electron spin-resolved and photon polarized ARPES experiment, with both the spin and orbital textures extracted in the same measurement. The predicted spin-orbital texture can be directly confirmed in this type of experiment, which can explicitly reveal how SOC plays a role in the real material at the atomic level.

\section{acknowledgments}
 We would like to thank Dr. Dan Dessau for sharing his experimental data and for useful discussions, which partially motivated this work. This work is supported by the Defense Advanced Research Projects Agency Microsystems Technology Office, MesoDynamic Architecture Program (MESO) through the contract number N66001-11-1-4105 and by the DARPA Program on "Topological Insulators -- Solid State Chemistry, New Materials and Properties", under the award number N66001-12-1-4034.

\end{document}